\documentclass[9pt, a4paper, sigconf, natbib=false, balance=false]{article}
\usepackage[utf8]{inputenc}
\usepackage{setspace}
\usepackage{dirtytalk}
\usepackage{float}
\usepackage[labelfont=bf]{caption}
\usepackage[section]{placeins}
\usepackage{multirow}
\usepackage{makecell, fancyhdr}
\usepackage{enumitem}
\usepackage{graphicx}
\usepackage{booktabs}
\usepackage{authblk}
\usepackage[english]{babel}

\usepackage[top=2cm,bottom=2cm,left=1.75cm,right=1.75cm,marginparwidth=1.75cm]{geometry}

\usepackage{multicol}
\usepackage{hyperref}
\hypersetup{colorlinks,allcolors=black}

\usepackage[backend=biber,style=numeric-comp, natbib=false]{biblatex}
\newenvironment{Figure}
  {\par\medskip\noindent\minipage{\linewidth}}
  {\endminipage\par\medskip}

\newenvironment{Table}
  {\par\medskip\noindent\minipage{\linewidth}}
  {\endminipage\par\medskip}

\title{Going public: the role of public participation approaches in commercial AI labs}
\date{\vspace{-5ex}}
\author{\href{http://orcid.org/1234-5678-9012}{Lara Groves}}
\author{Aidan Peppin}
\author{Andrew Strait}
\author{Jenny Brennan}
\affil{Ada Lovelace Institute, London, UK}

\begin{document}
\maketitle

\begin{multicols}{2}
{
\section*{Abstract}

In recent years, discussions of responsible AI practices have seen growing support for ‘participatory AI’ approaches, intended to involve members of the public in the design and development of AI systems. Prior research has identified a lack of standardised methods or approaches for how to use participatory approaches in the AI development process. At present, there is a dearth of evidence on attitudes to and approaches for participation in the sites driving major AI developments: commercial AI labs. Through 12 semi-structured interviews with industry practitioners and subject-matter experts, this paper explores how commercial AI labs understand participatory AI approaches and the obstacles they have faced implementing these practices in the development of AI systems and research. We find that while interviewees view participation as a normative project that helps achieve ‘societally beneficial’ AI systems, practitioners face numerous barriers to embedding participatory approaches in their companies: participation is expensive and resource intensive, it is ‘atomised’ within companies, there is concern about exploitation, there is no incentive to be transparent about its adoption, and it is complicated by a lack of clear context. These barriers result in a piecemeal approach to participation that confers no decision-making power to participants and has little ongoing impact for AI labs. This paper’s contribution is to provide novel empirical research on the implementation of public participation in commercial AI labs, and shed light on the current challenges of using participatory approaches in this context.

\section{Introduction}
Artificial intelligence research and technology continues to proliferate widely, presenting substantial opportunities but also considerable ethical risks for people and society. Against this backdrop, policymakers, researchers and practitioners are increasingly interested in public participation in AI: methods that enable members of the public to be involved and have their ideas, beliefs, and values integrated into the design and development process of AI systems \cite{berditchevskaia_participatory_2021, birhane_values_2022, lloyd_camera_2020}. 
There are two main reasons for this interest: the first is the perceived success of public participation and engagement methodologies in other fields: participatory approaches are used to address issues where there is impact on the public such as in international development \cite{cleaver_paradoxes_1999-1}, environmental justice \cite{glimmerveen_who_2022} and in democratic institutions \cite{dryzek_crisis_2019}. Increased interest in public participation in AI reflects a broader recognition of AI’s implications in the wider world. The second is the, by now, well-documented potential for AI systems to cause harm, such as causing discriminatory impacts on different members of society \cite{banerjee_reading_2022,buolamwini_gender_2018}, especially those from marginalised or disadvantaged backgrounds \cite{eubanks_automating_2018,kalluri_dont_2020-1}. Proponents of participation cite these methods as a way to create external scrutiny and accountability for these systems \cite{ada_lovelace_institute_algorithmic_2021,moss_assembling_2021}, and argue ‘more or better’ participation in AI \cite{himmelreich_against_2022} may partly remedy potential harms \cite{bondi_envisioning_2021,katell_toward_2020} and produce more ‘socially good’ outcomes \cite{bondi_envisioning_2021}.  Despite this growing interest, it is important to bear in mind that public participation is not a panacea for the harms that AI systems can raise, nor independently capable of deriving societal benefits of emerging technologies. Existing research around ‘participation washing’ highlights the potential pitfalls and extractive practices of these methods \cite{gilman_beyond_2022,sloane_participation_2020}. \par

A review of the literature at the interface between ‘participation’ and ‘AI’ reveals that, to date, there is very limited research exploring the role of public participation in commercial AI labs. There is also lingering conceptual confusion about what ‘participation’ in AI means and what kinds of approaches should be adopted \cite{birhane_power_2022,delgado_stakeholder_2021}, likely hindering wider adoption of these methods. Given that a significant proportion of AI development is undertaken in industry, there is a pressing need to understand how participation is, or could be, embedded in companies driving important developments in AI products and research. This need is all the more urgent in the context of the latest ‘AI spring’: the advent of novel general purpose and generative AI technologies, which may impact people at greater scale and in more unpredictable ways than traditional ‘narrow’ AI systems. Tech industry leaders have made calls for more ‘public input’ into systems like ChatGPT and GPT-4 to ensure these systems are aligned with societal needs \cite{openai_how_2023}. There have also been calls from industry leaders to ‘democratise AI’, a term that can have different or even conflicting meanings, such as increasing access to these systems or sharing governance of these systems \cite{seger_democratising_2023}. These developments have intensified the debate about what public participation in AI means.\par
This paper explores which public participation approaches are being used or considered by tech companies, how they understand the value of these methods, what barriers they face in using these approaches, and what impact public participation has on the company and on participants. Using a literature review of public participation in AI and 12 semi-structured interviews – nine with practitioners working at major AI-focused tech firms, three with non-industry professionals with a stake in the ongoing direction of ‘participatory AI’ –  conducted in the autumn of 2022, this paper seeks to answer three research questions:\par
\begin{enumerate}[label=\textbf{RQ\arabic*:}, align=left]
\item How do commercial AI labs understand public participation in the development of their products and research?
\item What approaches to public participation do commercial AI labs adopt? 
\item What obstacles/challenges do labs face when implementing these approaches?
\end{enumerate}

The contribution of this paper is twofold: novel empirical research reporting perspectives towards and past projects on public participation in commercial AI, and analysis on a current gap in the literature on ‘participatory AI’, finding that effective uses of participatory methods require a clear understanding of the context in which an AI system will be used. 

\section{Methodology}
Our findings emerge from two research verticals: a literature review and semi-structured expert interviews.

\subsection{Literature review}
We surveyed relevant literature on AI ethics and participation, the wider human-computer interaction (HCI), computer supported cooperative work (CSCW) and value-sensitive design (VSD) literature for scholarship on embedding participation in non-AI/ML technologies. We also drew on wider literature focused on the intersections of participation and democracy, for example, including deliberative democracy and sociology. We manually sourced literature from ACM and \textit{arXiv} repositories, using a combination of keyword searches: ‘public participation in AI’, ‘participatory AI’, ‘participatory design in AI’ and ‘public engagement’, as well as terms and concepts likely to yield discussion of similar/adjacent theoretical grounding including ‘social choice’ ‘and ‘democratising AI’. We also used a ‘snowball method’ to identify additional papers from reference lists.\par

\subsection{Expert interviews}

We conducted 12 semi-structured interviews in this research. The interviews were led by the lead author, with support and contributions from the second and fourth authors. We interviewed nine practitioners working in large, medium and start-up commercial AI labs developing both products and research, who may be involved in planning or implementation of public engagement / participation projects or be expected to carry forward findings of public participation projects into research and or product development. For additional background, we also interviewed three subject-matter experts across participatory design, participatory AI and public engagement methods, and with knowledge of tech industry practice. One of these three experts is employed by a technology-focused non-profit, two are currently employed by academic institutions; one of these two had recent previous employment in a commercial lab. All three have authored papers pertaining to participation in AI.  See Table \ref{tab:1} for participant IDs. Our interview questions were split into four sections. We asked participants:  
\begin{enumerate}
  \item How they understand public participation;
  \item What they think public participation in AI is for;
  \item What methods or approaches they have used in their work, or seen in use across the sector, and;
  \item Details of their role, their organisation's work culture, resources, and its propensity to fund or conduct participatory work
\end{enumerate}
\begin{Table}
\centering
\captionof{table}{\textbf{Participant organisation and ID}}
\label{tab:1}
\begin{tabular}{p{4.75cm}l} 
\toprule
\textbf{Organisation} & \textbf{Participant ID}\\
\midrule
    Start-up providing open source machine learning & P1 \\
    Large company developing both products and research & P2\\
    Large company developing both products and research & P3\\
    Large company developing both products and research & P4 \\
    Start-up developing research & P5\\
    Start-up providing open source machine learning & P6\\
    Company developing research & P7\\
    Tech-focused non-profit organisation & P8\\
    Academic institution & P9\\
    Academic institution & P10\\
    Start-up developing products (pre-market) & P11 \\
    Company developing research & P12\\
\bottomrule
\end{tabular}
\end{Table}
\par Participants were recruited either directly (selected based on previous demonstrable interest in ‘participatory AI’, ‘responsible AI’ or similar fields, and/or were part of the authors’ existing industry networks) and through snowball recruitment from recommendations from interviewees. Interviews lasted 60 minutes and took place virtually, using video conferencing software from September 2022 to January 2023, and were transcribed using a speech-to-text transcription software service. Three interviewees did not consent for their interview quotes to be used in this paper. Since all participants were in continuous employment at the time of participation, they were not offered additional payment for their time.
\subsection{Data analysis}
Interview data was analysed using a constructivist qualitative thematic analysis that draws heavily on a ‘theoretically flexible’  approach set out by Braun and Clarke (2006), that specialises in understanding and reporting repeated patterns, particularly in terms of institutional/organisational behaviours \cite{braun_using_2006}. 
Using a constructivist epistemology allowed us to approach the data with an understanding that meaning and experience are socially (re)produced 
\cite{burr_introduction_2006}. 
Following this paradigm, we coded our data and constructed our themes according to a ‘latent classification’ approach \cite{braun_using_2006} surfacing implied beliefs. \par
The interviews were coded by the lead author using data analysis software. We chose not to set prescriptive benchmarks around prevalence of codes, or whether codes directly related to the RQs. After an initial batch of 71 codes generated, a re-coding process resulted in 56: some codes were felt to be too broad, in other cases, two substantively similar codes were merged (e.g. ‘building rapport’ to ‘relationship building’), and antonyms such as ‘inclusion’ and ‘exclusion’ were felt to be usefully interpreted dialectically and coded as single entities. \par
From these 56 codes, reproduced across Tables \ref{tab:2} and \ref{tab:4}, we identified six main themes that corresponded to different research questions:
\begin{enumerate}
    \item Internal factors
    \item Commercial factors
    \item Field-level factors 
    \item Societal and moral factors 
    \item Purpose of participation
    \item Participatory approaches
\end{enumerate}
From the data, we surfaced many different operational considerations and personal values/beliefs that practitioners suggested are (or might be) impactful for the adoption of public participation. Factors were reported to emanate from the level of the firm (‘Internal’), or externally (‘Field-level’), and pertained to business mission (‘Commercial’) or relationship to people and society (‘Societal and moral’). These are categorised as ‘factors’ over the more directional e.g. ‘blockers’ or ‘drivers’ to avoid setting up a simplistic binary for phenomena not experienced by all participants universally. Some codes appear in different themes, highlighting the porous boundaries between these themes. Theme 5 and Theme 6 concern methods and approaches for, and purpose of, participation, and therefore correspond explicitly with RQ1 and RQ2 of our study.\par

\begin{Table}
\centering
\captionof{table}{\textbf{Themes and codes constructed from factors relevant to the adoption of public participation in commercial AI (as reported by interviewees)}}
\label{tab:2}
\begin{tabular}{p{2.5cm}p{5cm}}
\toprule
\textbf{Themes} & \textbf{Codes} \\
\midrule
\multirow{8}{2.5cm}{Internal factors} & Buy-in for public participation \\
    & Compensating participants \\
    & Internal expertise \\
    & Remit: AI product or AI research \\
    & Responsibility for public participation \\
    & Scale and scope of public participation \\
    & Types of `public' \\
    & Capacity building \\
\midrule
\multirow{3}{2.5cm}{Commercial factors} & Profit motive \\
    & PR, optics, reputation \\
    & Transparency \\
\midrule
\multirow{6}{2.5cm}{Field-level factors} & Capacity building \\
    & Intermediaries \\
    & Lack of industry-specific methods or training on public participation \\
    & PR, optics, reputation \\
    & Regulation \\
    & Responsibility for public participation \\
\midrule
\multirow{8}{2.5cm}{Societal and moral factors} & Extractive practice \\
    & Good intent, social good \\
    & Harms, discrimination \\
    & (In)justice, (in)equality \\
    & Inclusion, exclusion \\
    & Power \\
    & Society building \\
    & Trustworthiness \\
\midrule
\multirow{10}{2.5cm}{Purpose of participation} & Democratising AI \\
    & Good intent, social good \\
    & Good business \\
    & Widening inclusion \\
    & Embedding lived experience \\
    & Intrinsic value of participation \\
    & Public participation as a form of accountability \\
    & Relationship building \\
    & Soliciting input / knowledge transfer \\
    & Trust building \\
\bottomrule
\end{tabular}
\end{Table}

\begin{Table}
\centering
\captionof*{table}{\textbf{Table 2 cont. - Themes and codes constructed from factors relevant to the adoption of public participation in commercial AI (as reported by interviewees).}}
\begin{tabular}{p{2.5cm}p{5cm}}
\toprule
{\textbf{Themes}} & \textbf{Codes}\\
\midrule
& Citizens' jury\\ 
& Crowdsourcing\\
& Co-design\\
& Community training in AI\\
& Community-based approaches\\
& Community-based Systems Dynamics framework\\
& Consultation\\
\multirow{4}{*}{\makecell[cc]{Participatory\\ approaches}} & Cooperatives\\
& Deliberative approaches\\
& Diverse Voices method\\
& Fairness checklist\\
& Governance tools e.g. audits, impact assessments, other policy mechanisms\\
& Open source\\
& Participatory design\\
& Request for comment\\
& Speculative design/anticipatory futures\\
& Surveys\\
& User research/user testing\\
& Workshops/convenings\\
\bottomrule
\end{tabular}
\end{Table}
\subsection{Positionality statement}
At the time of research, all the authors were employed by an independent research institute that conducts evidence-based research on data and AI in policy and practice, with a core organisational belief that benefits of data and AI must be justly and equitably distributed, and must enhance individual and social wellbeing. As part of the organisational remit, the institute collaborates with technology companies in a research capacity, i.e. using industry as a site of study. It does not accept funding from technology companies. The authors live and reside in the UK, and two of the four authors are British, one is British and Irish and one is American. We adopt a sociotechnical conception of AI, understanding that the technical elements of AI – machine learning, neural networks, etc – are inherently interrelated with social, political and cultural factors, principles and motivations (see for example Mohamed et al.) \cite{mohamed_decolonial_2020}.

\section{Literature review}

\subsection{Public participation in theory and practice }
Broadly in the literature, public participation refers to approaches or activities that engage or involve members of the public, incorporating perspectives and experience into a project or intervention.  Participatory approaches are routinely adopted in a number of areas, environmental decision-making \cite{glucker_public_2013, hugel_public_2020}, health and care \cite{ocloo_exploring_2017, russell_impact_2020} and in democratic institutions \cite{bloomfield_deliberation_2001, berditchevskaia_participatory_2021}. For example, feedback sessions in health and social care incorporate patient views and lived experience to inform ongoing service delivery (described as ‘patient and public involvement’ (PPI) in the UK) \cite{bit_deliberative_2022} and consultations in policy mechanisms such as environmental impact assessments foster democratic debate and broaden decision-making powers \cite{glucker_public_2013}.\par
In technology design contexts, participatory approaches stem from the fields of human-computer interaction (HCI) \cite{jacko_human-computer_2003}, user-centred design \cite{lloyd_camera_2020} and the theory and application of participatory design (PD) methods \cite{pierre_getting_2021}. These fields offer critical examination of how design might be crafted in tandem with \cite{harrington_forgotten_2020}, instead of on behalf of, different publics in order to incorporate their needs and values \cite{russell_impact_2020, sloane_make_2022, habermas_between_1996, agid_its_2016,bratteteig_unpacking_2016}. In deliberative democratic theory, it is argued public participation appeals to democratic ideals of legitimacy \cite{solomon_why_2012-1} and accountability \cite{bovens_analysing_2007-1} as well as to enhance political autonomy \cite{habermas_between_1996}. The tradition of deliberative participation – the involvement of the public with a view to fostering deliberative debate and engagement – is evident in participatory design,  which offers participants ‘seats at the table’  \cite{probiner_smart_2018}, emulates democratic decision-making \cite{delgado_stakeholder_2021},  adopts consideration of social and political contexts \cite{agid_its_2016} and embraces co-production \cite{holstein_co-designing_2019}. Participation is also often read as an intrinsic value in and of itself \cite{flanders_what_2013}: like similar concepts such as ‘inclusion’ or ‘collaboration’, it is often understood in the literature as indicative of a ‘moral good’ \cite{himmelreich_against_2022}, of ‘flourishing social ties’\cite{brodie_understanding_2009} and so on. However, within the literature, there is little agreement about who constitutes the ‘public’. In politics and policy domains, the ‘public’ may refer to ‘citizens’, ‘labelling data people’ or ‘laypersons’ \cite{glimmerveen_who_2022} while, in technology contexts, it may refer to current or future ‘end users’ \cite{park_ai-based_2019}. More recent literature around participation in AI adopts a broader definition that includes all people affected by the use of an AI system, particularly individuals and groups for whom AI risks exacerbating inequity, injustice and marginalisation \cite{robertson_what_2020}. This raises the question of how commercial AI labs define ‘public’ in any public participation activities, particularly when their technologies may impact multiple publics in multiple areas or regions. \par
The form of public participation can vary, reflected in the various typologies produced by political scholars and practitioners \cite{iap2_core_nodate,cornwall_unpacking_2008}. The first of these is Sherry Arnstein’s Ladder of Citizen Participation \cite{arnstein_ladder_1969}, a widely referenced framework for forms of participation, originally intended to outline different degrees of participatory approaches in public planning. Arnstein’s eight rungs range from forms of non-participation (‘manipulation’), one-way dialogic methods (such as public request for comment \cite{kochan_commenting_2017}), involvement by consultation and partnership in the middle rungs, and finally ‘citizen control’ at the top rung (see Figure \ref{fig:my_label1}). Arnstein is critical of approaches at the bottom of the ladder, branding them tokenistic and inadequate in shifting the axis of power and therefore not paramount to meaningful participation \cite{arnstein_ladder_1969, birhane_power_2022}. \par
\begin{Figure}
\centering
\includegraphics[width=\linewidth]{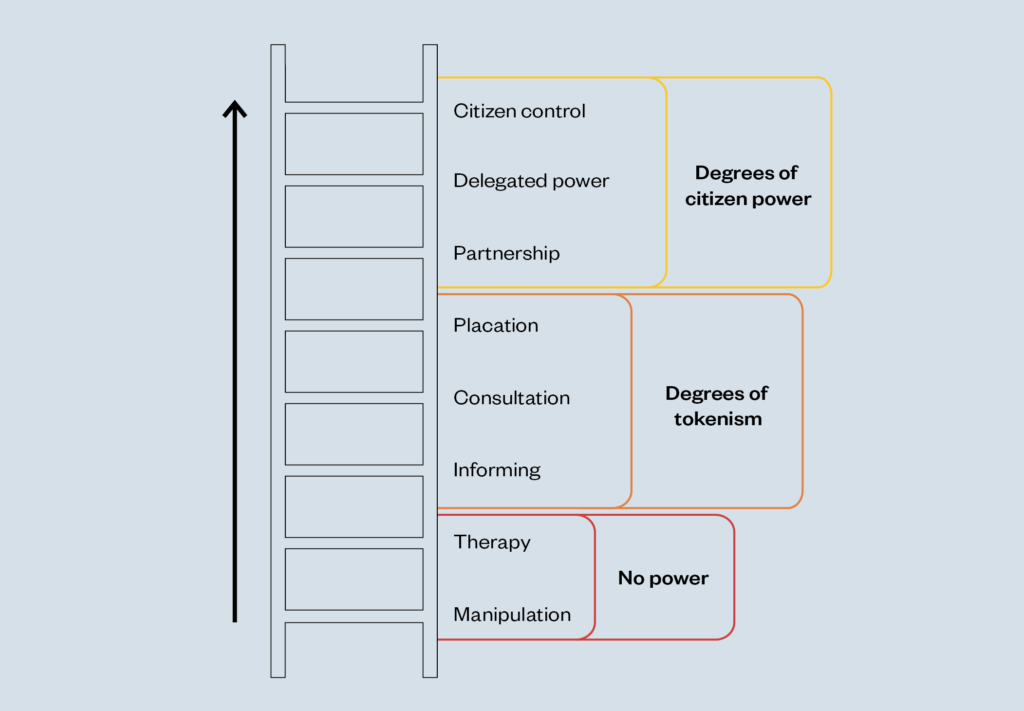}
\captionof{figure}{\textbf{Arnstein's `Ladder of Citizen Participation' \protect\cite{patel_participatory_2021} \protect\cite{arnstein_ladder_1969}}}
\label{fig:my_label1}
\end{Figure}
\begin{Figure}
\centering
\includegraphics[width=\linewidth]{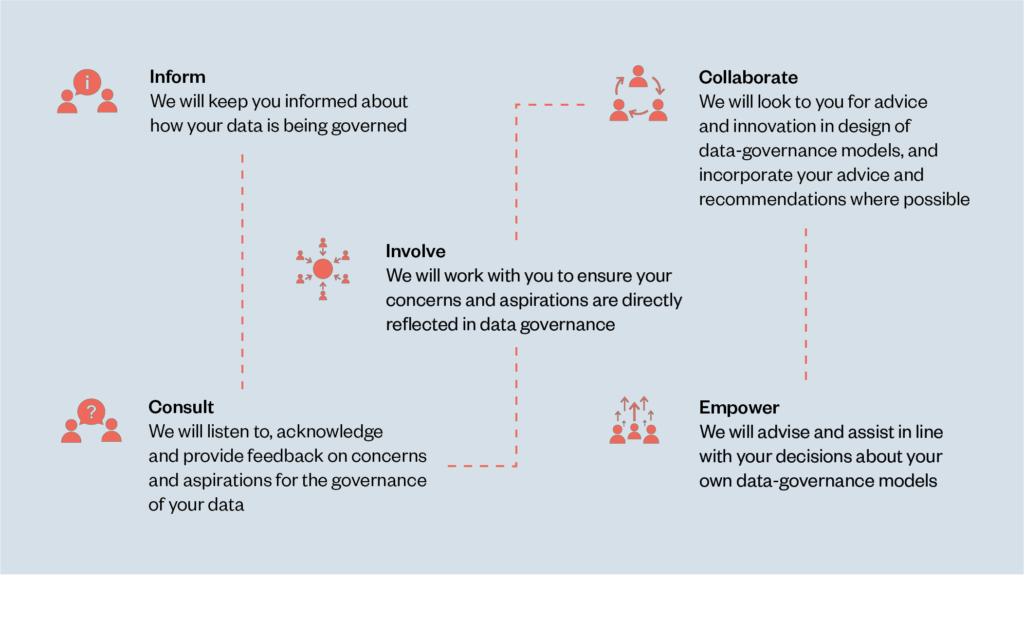}
\captionof{figure}{\textbf{Framework for Participatory Data Stewardship \protect\cite{patel_participatory_2021}}}
\label{fig:my_label2}
\end{Figure}
Patel et al. \cite{patel_participatory_2021} draw on Arnstein’s ladder and a more recent ‘spectrum of participation’ \cite{iap2_core_nodate} to describe practical mechanisms of participation in the stewardship of data and consequently the design of data-driven systems, including AI (see Figure \ref{fig:my_label2}). Their analysis creates a link between Arnstein’s political lens on participation and participation in sociotechnical contexts by describing five levels of participation and examples of what practical mechanisms may exist for each, drawn from real-world case studies. These five levels include:
\begin{enumerate}
    \item Informing people about how data about them is used, such as through the publication of model cards;
    \item Consulting people to understand their needs and concerns in relation to data use, such as through user experience research or consumer surveys;
   \item Involving people in the governance of data, such as through public deliberation or lived experience panels;
    \item Collaborating with people in the design of data governance structures and the technologies they relate to, such as through novel institutional structures like ‘data trusts’, and;
    \item Empowering people to make decisions about datasets and technologies built with them, such as through citizen-led governance boards.
\end{enumerate}
Though indirectly linked to AI, these taxonomies help us makes sense of the public participation approaches commercial AI labs may be using and contribute theoretical foundational frameworks for exploring participation in AI design.

\subsection{Public participation in AI technology development}

As Dove et al. note, AI is neither ‘arcane nor obscure’ \cite{dove_ux_2017}: discursive debate around participation in AI should not be isolated from debates around participation more generally. Cooper et al. argue the AI design and development pipeline of AI technologies is diffuse and therefore typically ‘participatory’, combining multiple iterative activities and the input of multiple actors \cite{cooper_accountability_2022} across 'algorithmic supply chains' \cite{cobbe_understanding_2023}.  However, as with participation adopted in other domains, there are varying possible degrees of participation in AI. 
Two existing typologies are instructive for classifying the different modes of participation in AI: Sloane et al.’s typology of participation: \textit{as work, as consultation} and \textit{as justice} \cite{sloane_participation_2020}, and Birhane et al.’s exploration of the three instrumental categories of participation: \textit{for algorithmic performance improvement}; \textit{for process improvemen}t and \textit{for collective exploration} \cite{birhane_power_2022}. These typologies provide a sense of some of the goals of public participation in AI, and where participatory approaches can fit in AI development or research. \par
There is an emerging literature on participatory approaches to AI development, which identify a few kinds of ‘participatory’ activities that involve assembling a mixed group of stakeholders to consult or assess an AI system. The literature on participatory development highlights a few activities that are seen as ‘participatory’. These include crowdsourcing \cite{diaz_crowdworksheets_2022,vaughan_making_2017} (such as crowdsourcing possible impacts of ADM systems \cite{barnett_crowdsourcing_2022} or labelling data \cite{park_ai-based_2019}), participatory dataset documentation \cite{suresh_towards_2022}, creating ‘red teams’ to test or evaluate a model \cite{ganguli_red_2022}, bug bounties \cite{openai_announcing_2023} or engaging members of the public to elicit preferences for  algorithmic design decisions \cite{christiano_deep_2017, robertson_what_2020}. Such forms of participation very often prioritise a higher total number of participants over length or depth of participant involvement \cite{barnett_crowdsourcing_2022}. For example, participatory development of ML datasets \cite{denton_bringing_2020}, requiring higher degrees of input from a higher number of stakeholders might be classified as Sloane’s ‘participation as work’, where methods that foster  deliberation  around values and experience \cite{dilhac_responsible_2021}, might fall under Birhane et al.’s heading of ‘collective exploration’. Other scholarship argues that participatory approaches in AI could be instrumentalised to advance ambitious societal-level goals such as fairness, inclusion \cite{frey_artificial_2020,stilgoe_developing_2013}, justice \cite{costanza-chock_design_2020, scott_algorithmic_2022}, accountability \cite{boag_tech_2022} and democratic values \cite{frey_this_2019}, which could be characterised as Sloane et al.’s ‘participation as justice’ \cite{sloane_participation_2020}. 
Birhane et al. offer three case studies of a participatory approach to AI development, instances where participation is sought to improve the function of large language models for African and Te Reo Māori languages, annotate datasets and improve dataset documentation \cite{birhane_power_2022}. The authors suggest community inclusion in such projects might advance goals such as equity and justice, but acknowledge that participation in these kinds of projects may amount to products built that actually harm the communities included. Another proposed method for participation in AI development is Martin, Jr. et al.’s Community Based System Dynamics (CBSD) method, a mechanism that seeks to ‘engage and centre perspectives of marginalized and vulnerable communities’ for the purposes of model refinement \cite{martin_jr_participatory_2020}, however only offering cursory detail on the methodological components required to achieve this goal. \par
There are concerns of ‘participation washing’ \cite{sloane_participation_2020} across participation literature, also highlighted in application to AI. Hossain and Ahmed note that, to date, participation in design or development of AI has been overly modest and inconsequential, prescribing only narrow technological solutions as opposed to lasting community or societal change \cite{hossain_towards_2021}, following the general mode of critique from the participation literature \cite{cleaver_paradoxes_1999-1,glimmerveen_who_2022}. Sloane et al. argue that participatory approaches that claim to value diverse expertise and express a commitment to recentring marginalised communities, but in practice function as (often unrecognised) labour, risk paying lip service to the pro-social ends of participation while exploiting disadvantaged groups \cite{couldry_decolonial_2021,sloane_participation_2020}. There are also dangers, as noted by Lloyd et al., that with a focus on engaging technology ‘users’ (in participatory projects), users become a stand-in merely for ‘consumers’, narrowing focus away from broader segments of society that might be affected by AI, with a risk of exacerbating existing harms to these groups \cite{lloyd_camera_2020}. In instances where a wider focal point is adopted to target ‘non-users’ of technologies, often under the objective of ‘democratising AI’ \cite{dempsey_access_2021}, the outcome may not be equivalent to entrenching participatory or democratic structures \cite{saetra_avoid_2022} but may simply indicate intent to ‘widening access’ to technology use or development \cite{seger_democratising_2023}.
\subsection{Public participation in commercial AI}
Over the past decade, many large technology companies have established or acquired their own dedicated AI labs for developing research and products: for example, the AI research company DeepMind was acquired by Google in 2014 and is now a subsidiary of Google's parent company Alphabet. Google itself has invested in entire AI research wings like Google Brain, and has integrated AI research into its products. There are also a number of smaller, independent companies developing AI that have made significant research and product developments, such as OpenAI and their ChatGPT model and interface. Commercial AI labs are widely considered to be at the forefront of current AI development and research \cite{hogarth_state_2022}. \par
Many AI labs have teams that are specialised in ethics issues (Microsoft’s Office for Responsible AI, Google DeepMind’s Ethics and Society team), including a remit for activities such as public participation. Though debates around ethics, fairness and accountability have gained considerable traction in recent years, it is still challenging terrain: Moss and Metcalf point to a habitual inability among firms to to specifically designate which team (members) have the responsibility for embedding ethics \cite{moss_ethics_2020}, as well as an ineptitude toward institutionally buttressing their role(s), creating pinch points and barriers to the effective implementation of AI ethics initiatives. Practitioners struggle with what Rakova et al. identify as a demanding interplay between ‘organizational structures and algorithmic responsibility efforts’ \cite{rakova_where_2021}.  Other scholars have criticised tech companies have for ‘ethics washing’ behaviours,  \cite{bietti_ethics_2020} including the use of internal ethics initiatives as a form of social capital that justifies deregulation of their industry in favour of self regulation. \par
Despite the sheer quantity of industry-led AI/ML research, most scholarship on participation in AI to date has emanated from academia or civil society: there is scant publicly available evidence of what kinds of participatory methods or projects are put into use in commercial AI labs. What literature does exist on public participation approaches in industry is authored by individuals working in commercial AI ethics teams \cite{birhane_power_2022, martin_jr_participatory_2020}, and the limited examples we have of participatory efforts are also led by ethics teams in these companies. Examples include the Royal Society of Arts (RSA) and Google DeepMind’s Forum for Ethical AI project, involving a citizens’ jury with members of the public to offer space for deliberation on algorithmic decision-making \cite{the_rsa_democratising_2019}, and Behavioural Insights (BIT)’s blog on a recent partnership with Meta constructing citizens' assemblies for members of the public to deliberate on climate misinformation \cite{bit_deliberative_2022}. The lack of public examples of AI labs using participatory methods raises questions about the real extent of their use.

\section{Interview Findings} 
Based on our review of the literature, we asked our interview subjects how commercial AI labs understand participatory AI approaches and the obstacles they have faced implementing these practices in the development of AI systems:
\begin{enumerate}
\item Within commercial AI labs, public participation is viewed as serving societally `good' ends, but may also have a strong business purpose;
\item Public participation in AI industry lacks clear and shared understanding of practices. Participants did not identify many participatory methods they use, but rather tended to list methods they had heard of;
\item Public participation in AI labs faces various obstacles: resource-intensity, atomisation, exploitation risk and misaligned incentives;
\item Public participation in AI labs is complicated by products or research that lack a clear context.
\end{enumerate}
\subsection{Within commercial AI labs, public participation is viewed as serving societally ‘good’ ends, but may also be good for business}

\say{\textit{We do a lot of AI for social good projects at} [large company].\textit{ But I’m always wondering why we need the qualifier of AI for social good}.}  [P3]

Interviewees, including the practitioners working on ‘participatory AI’ and adjacent topics, view participation and participatory approaches positively, with several associating these practices with  ‘doing good in the world’, an indication of company legitimacy or as a commitment to accountability. Another participant described the pull to embed participatory approaches as an ‘obligation’, to ensure the company are achieving societally beneficial outcomes with their technologies:
\par
 \say{\textit{We, as a corporation, building or researching a technology that has the potential to solve problems for people, have an obligation to engage folks from various backgrounds to help us understand the different problems they face.}} [P7] \par 
Some interviewees report viewing public participation in the labs through the lens of profitability or business mission:  
\say{\textit{It should be for good business, right? Engaging with people should help you build a product that addresses their wants and needs better which in turn, makes your company more profitable.}} [P2] \par 
This view more closely follows the argument that increasing participation in corporate tech contexts presents an opportunity to increase access to technology: unsurprisingly, if your goal is to build better tech, then making it work better for more people is an attractive prospect. However, other participants expressed frustration that this would be likely to be the only logic that would wash with corporate shareholders (who, as one interviewee suggested, would not find any reason to complain if public participation was not conducted at all). In larger companies, interviewees noted challenges of explaining the value and role that participation can play to others in the firm. Those using these methods were trying to resolve concerns around, for example, bias and fairness, but often found that they had to reframe these objectives from the perspectives of how these methods could provide an increasing return on revenue. One interview noted concerns of a performativity around labels such as ‘responsible AI’: \par
\say{\textit{There’s concern about being exploitative in using that knowledge to do this sort of marketing veneer of responsible AI, then we’re still just going to make money on everything}.} [P3] 

\subsection{Public participation in commercial AI lacks a clear and shared understanding of practices}
   \say{\textit{We take that there are many different approaches to public participation } [at the company]\textit{. Some are more kind of focused on participatory annotation of data and co-production of AI systems. I think my work is more focused on vision setting for the future of AI.}} [P7] \par 
   Our research corroborates findings from the literature of an enduring lack of consensus around participatory approaches in practice \cite{birhane_power_2022,costanza-chock_design_2020}. Interviewees were asked ``what approaches to participation have you used in your work or practice?" Some interviewees were able to talk about approaches they’d personally used for certain research/development projects, but more usually, would recall (often cursory) detail about specific projects or ideas in either their organisation or across the sector, rather than any direct experience. Overall, interviewees cited 1 different methods they were familiar with or had used – see Table \ref{tab:4}. 
\begin{Table}
\centering
\captionof{table}{\textbf{Participatory approaches in commercial AI (as reported by interviewees) mapped onto Arnstein's `Ladder of Citizen Participation'}}
\label{tab:4}
\begin{tabular}{p{2.5cm}p{4.75cm}}
\toprule
\textbf{Arnstein's ladder} & \textbf{Participatory approaches} \\
\midrule
\multirow{8}{2.5cm}{Degrees of citizen power} & Cooperatives \\
    & Citizens' jury \\
    & Community-based approaches \\
    & Deliberative approaches \\
    & Participatory design\\
    & Speculative design/ \\
    & anticipatory futures\\
    & Governance tools e.g. audits, \\
    & impact assessments, other policy mechanisms\\
\midrule
\multirow{8}{2.5cm}{Degrees of tokenism} & Co-design \\
    & Community training in AI \\
    & Community-based Systems Dynamics \\
    & framework \\
    & Crowdsourcing \\
    &  UX/user testing \\
    &  Open source \\
    &  Diverse Voices method\\
    & Workshops/convenings \\
    & Consultation \\
\midrule
\multirow{2}{2.5cm}{Non participation} & Surveys \\
    & Request for comment \\
\bottomrule
\end{tabular}
\end{Table}

The method interviewees cited  most often was a form of consultation with people outside the company, generally domain experts rather than members of the public, usually to solicit feedback on the design or usability of products. Most interviewees recognised that participation could have multiple dimensions, with a few specifically using the word ‘spectrum’. 
Two interviewees suggested that open sourcing machine learning models, as a kind of mass participation predicated on widespread involvement, might constitute a participatory approach. \par
Despite overall understanding and knowledge of types of approaches that could be used in AI development, the important accompanying finding is that most interviewees did not feel fully equipped to report on their organisation’s activity in the area of ‘participatory AI’. While we cannot rule out that commercial AI labs are using participatory methods that we are unaware of, these findings suggest that, at best, interviewees did not feel comfortable discussing specific examples of these methods with us, or had no awareness of these methods being used in their companies – and at worst, that such methods are not being used at all. Given that most interviewees self-selected to participate on the basis of their familiarity with public participation in AI (see ‘Methodology’), it would appear that the most likely scenario is that there is little use of participatory approaches to AI in industry.
\subsection{Public participation in commercial AI labs faces various obstacles: resource intensity, atomisation, exploitation risk and misaligned incentives}
\subsubsection{Embedding participation is expensive and resource-intensive} 
\say{\textit{If you want actual participation, you actually have to invest before you need something from people.}} [P2]\par
As reported elsewhere in the literature \cite{delgado_stakeholder_2021}, practitioners we spoke with struggled to embed participation in their companies. The accordant time and costs, and the difficulty in quantifying the work, are at present seen as too great to inspire action (and therefore outweighing any motivations for ‘social good’). One interviewee put forward interest in conducting further participatory work, but felt that other research and development pursuits, like ensuring ‘truthfulness’ of large language models, would be a higher priority. Many interviewees put forward a need for capacity building in this space, stating that, at present, practitioners are not equipped to conduct public participation, as many do not come from a social science background or have not undertaken work with community groups, and therefore lack the requisite skills and experience to undertake long-term engagements with members of the public.\par

\subsubsection{Participation in the AI industry is ‘atomised’}  
Interviewees often expressed there was not a clear understanding within AI companies of who has the responsibility for leading participatory projects or embedding a ‘culture of participation’ in which all members of a product team have a shared understanding of the value and uses of these methods. One interviewee suggested that spearheading the adoption of public participation in AI labs puts you at odds with the direction of travel of the rest of the company, effectively creating misaligned incentives, with public participation work not rewarded or recognised within the organisation. In the cases our interviewees mentioned, participation generally arose emergently, responding to specific design or development knots (particularly in the ‘agile development’ \cite{delgado_stakeholder_2021} of product lifecycles). One interviewee pointed to burnout and a lack of bandwidth among tech workers, preventing individual practitioners from connecting with other individuals or teams who had taken on participatory work in the past. \par

\subsubsection{There is concern and care around exploitation and ‘participation washing’} 
Many interviewees report that they are paying attention to social, societal and moral questions when considering how to adopt public participation approaches in their practice. Frequently cited considerations include concern about extractive behaviour and practice, whether or not ‘inclusion’ is always a commendable value, and questions of power, justice and societal impacts. Two interviewees specifically cited the term ‘participation washing’ \cite{sloane_participation_2020} when sharing thoughts on potential obstacles to embedding participation, which may indicate that this is a concern that has become more routinely observed in these companies.  
Most interviewees reported feeling great responsibility for non-tokenistic participation and being attuned to power and privilege, especially in capacity as a tech worker. While these interviewees demonstrate a motivation for wanting to adopt meaningful participation that confers decision-making power for participants, for many, it did not translate into ‘better’ participation (often because owing to the other obstacles we set out in this paper, they felt they could not do a deeper level of engagement justice). Some interview subjects highlighted the tension between the business needs of a commercial lab and the mode of participation in certain projects.
While one interviewee reported satisfactory levels of funding and support received by their company, this puts undue pressure on wanting to achieve the ‘desired’ outcomes from participatory work, recalling a project where they were told to \say{\textit{go back and get a different answer} [from participants]}  [P3]. 
Other interviewees described concerns of exploitation of participants from marginalised or underrepresented communities in their work:\par
[recalling previous public participation in the company]\say{\textit{It gets to the point where it’s like}  ‘\textit{Oh, yeah, we talked to some Black people. And they said it’s fine.’ And we’re being fair! We’re being responsible!}}  [P3] \par
Practitioners report grappling with values such as societal justice and the relation to their work: some discussion across different interviews took place on whether ‘inclusion’ in AI could advance justice or address power asymmetries. Most interviewees were firm on the importance of adopting focus on communities that have historically been excluded from technology development conversations. For some companies, lowering the barrier of participation/inclusion in AI was deemed a priority, usually in the context of enabling different groups of people to design or use machine learning tools. Moreover, some interviewees situated the role of participation into the broader societal context: one participant argued the role of participation is interrelated to broader questions of political representation and governance:
\say{\textit{That’s the realm of the political, setting up the terms under which we all live together. And increasingly, technology, technology systems have encroached so thoroughly on that, that we're having to rethink all of these extremely old questions about how can people self-determine the conditions under which they live in a technology space?}} [P10] \par
Concern and care over extractive practice and exploitation was reported to closely correlate with the type of ‘public’ chosen to take part in participatory projects: two interviewees revealed that it is often subject-matter experts that are assembled in place of ‘laypeople’, suggesting that technical expertise is more often sought out by companies than lived experience. This echoes concerns in the literature around which publics are participating, a particular concern for public participation in AI given the potential for AI systems to impact communities across the globe at great scale and magnitude. \par

\subsubsection{Commercial AI labs are not incentivised to be transparent or share their experiences using participatory approaches}  
Even where participatory approaches are tested and trialled, interviewees described a lack of incentive to report publicly about the work and any potential learnings. One suggested that publishing detail on participatory approaches and specific methodological choices might pose a commercial risk, as it would be sharing information that could be seen as intellectual property.\par
Some interviewees reported feeling conscious about the reputation of their company, and the ways in which publicising (or not publicising) certain activities could be seen as affecting optics and comprising good or bad ‘PR’, suggesting that this disincentivises experimenting with public participation.\par
One interviewee reported feeling as though external scrutiny over practice and public pressure to enact their social responsibilities (where they saw participatory work as situated) did not have much of an effect on the company’s direction or bottom line at all:  
\say{\textit{If you take all the headlines} [on tech industry practice] \textit{over the last five years, they didn't affect share price, or revenue}} [P3] \par 
This suggests that, for this company, ‘the techlash’ \cite{atkinson_policymakers_2019} has not had enormous impact on their practices and would not incentivise publishing details of participatory approaches. \par
A lack of transparency has effects at the industry level. Institutional theory holds that companies in the sector begin to homogenise when faced with the same set of economic conditions \cite{dimaggio_iron_1983}, and one interviewee reported that this felt true of  tech companies – \say{\textit{all the AI companies just look at each other} }[P3], suggesting a ‘fear of missing out’ effect. 
 Coordinated, tech industry-wide effort was often cited by interviewees as being critical for an ecosystem of public participation, particularly around pooling resources to collectively establish or articulate better participatory practices. Most interviewees saw an increased role for some kind of regulation to incentivise public participation, though not without caveat:   
\say{\textit{That’s a whole other issue of “gaming” regulation. You know, you start this cat and mouse game of: “Here’s some regulations”. And then companies are thinking, how do we get out of this?}}[P3] \par
Other actors’ contributions to deriving change across the sector was noted by some interviewees, particularly activists. Some suggested looking to other sectors to use as analogues for an AI industry-specific approach. The FDA’s medical device pipeline, with its requirement for patient involvement, was offered by two interviewees in this context, as a potential practice that could be adapted to AI research and development. \par

\subsubsection{Participation in commercial AI labs is complicated by products or research that lack clear context} 
As demonstrated above, public participation is costly and resource intensive: companies already lack incentives to conduct it, and where it is conducted, it can be piecemeal. The difficulty of running public participation methods is exacerbated as the generalisability of AI increases. \par
Three interviewees identified a need to conduct public participation work around more complex, general purpose AI systems where the context in which it could be used to impact the public is less clear, and an additional two were concerned about conducting public participation in the face of rapid development of general purpose AI systems that may present complexities for a non-technical ‘public’. \par
The interviewees we spoke with who belong to or work closely with AI product teams regularly conduct UX/user-research to get feedback on the usability of the proposed product with a narrow group of potential users. Interviewees saw this context as favourable for public input, as potential participants may have a clearer understanding of the impacts of the proposed system:  
\say{\textit{Being in a product team can be really focusing, because we have these goals for the conversation. So you can get much clearer feedback from } [participants]\textit{}} [P2] \par
One interviewee recalled a project assembling members of the public to discuss potential benefits, harms and use cases of AI models at a high level, but reported that the exercise lacked focus and was not perceived by their company to have useful impact. They suggested that using specific technologies as a steer might enable critical dialogue on possible societal impacts of a technology at a higher level (though did not feel well-equipped to conduct such approaches at present). \par
Interviewees belonging to research teams, outside strict product deadlines, put forward that they have more flexibility to pursue alternate research or design agendas. For example, practitioners working in research teams had encountered more methods akin to co-design \cite{hossain_towards_2021,lloyd_camera_2020} as a result of more agency to set pace and objectives. We find that embedding far-reaching or longer-term public participation projects is seen as particularly complex for general purpose technologies that have many number of downstream applications. One interviewee expressed concern at the pace and spread of recent developments in generative AI further implicating the scope and scale of participation, as well as participant understanding:  
\say{\textit{What does it mean to engage people who are affected by, but don’t have the knowledge of, state of the art systems, especially as things like DALL-E and DALL-E mini} [now Craiyon] \textit{and Stable Diffusion go viral}?} [P1]\par
As generative AI and similar technologies continue to proliferate at an astonishing rate, with innumerable downstream uses and a wide user base, several interviewees reported the obligation to conduct some kind of public participation work across a variety of conditions increases, as highlighted by this quote:  
\say{\textit{The people that put} [content such as images] \textit{into the public sphere did not know they would be used for this application. How could you know that something you posted in 2007 would be used in a model over a decade later? So the public should have a say.}} [P2]  \par
These findings show that any proposed public participation approach or project must be attuned to the specific context of AI development (product or research).
Our findings reveal that it’s harder to do public participation when the context in which it would be used or affect the public is less clear (for example, in AI research that is theoretical rather than practical, or with AI systems like generative models that can impact or be used in multiple contexts relevant to a person's life).
\section{Limitations}
\subsection{Limitations of interview approach}
We report the following limitations of our interview approach:
\begin{itemize}
    \item[--] \textbf{Non-representative sample:} Not every major AI lab is represented in this study.  In the largest companies, we would have preferred to interview multiple employees from different teams to gain a richer understanding of institutional culture and practice, which is hard to glean from a single interview. Additionally, interviewees in many cases were selected (or self-selected) on the basis of pre-existing interest in ethical/participatory AI etc.
    \item[--] \textbf{Barriers to participation:} We identify two main barriers to participation: interviewee concern around candour, and atomisation of public participation in commercial environments.
\end{itemize}

\noindent Drawing from the research team's prior experience working in industry, and our experiences engaging with industry representatives, we recognised the potential for interviews to surface commercially sensitive IP and or corporate malpractice, resulting in varying degrees of comfort and willingness to interview. Many interviewees may have been reticent to share identifiable details of relevant projects within interview. While we sought to address this limitation by offering interviewees anonymisation of findings and removal of identifiable material, this concern may have persisted. Additionally, as we set out in the Discussion, there is often limited awareness both internally and externally on which individual/team has remit or expertise for public participation, arising in confusion over who would be best placed to participate in this study.

In total, 47 direct personal invitations were sent for this study, in addition to two broadcast messages on two ‘responsible tech’ Slack boards. 12 directly invited interviewees explicitly declined the offer of participation in this study, we speculate in part owing to some of the barriers set out above, in addition to burnout (which was explicitly cited by a couple of invitees). This resulted in a relatively small sample size of remaining respondents who were available and happy to interview.

\subsection{Limitations of study}
We acknowledge here the recent rounds of tech sector layoffs and the gloomier economic climate beginning to intensify during and shortly following our interview period, and suggest these will have tangible implications for the adoption of participatory approaches (but which are not specifically reported on or studied here). We are employed by a research institute operating in the UK and in Europe, and all interviewees are employed at companies or institutions located in North America and Europe, reflecting the dominant geographies of high-profile AI research labs. We would have preferred to have substantive input from organisations based in the Global Majority represented in this research, though we note, following Chan et al. and others, that mere inclusion is not a conduit to rebalancing North American power domination \cite{chan_limits_2021}. Nevertheless, there may be opportunity for future research along these lines.

\section{Conclusion}
In this study, we find that although public participation is recognised as a valuable mechanism to involve public perspectives and enjoys support and interest from this sample of interviewees in commercial AI labs, only limited participatory projects have been explored and implemented to date. Commercial AI labs view public participation as a way to mitigate ethical risks in AI systems and produce more ‘societally beneficial’ technologies. However, our interviewees report that individuals responsible for implementing participatory approaches in commercial labs do not have a shared understanding of what methods can or should be used and how to use them. While many of the challenges of embedding public participation are not unique to the commercial sector, nor to the context of technology development, there are routinely observed difficulties for public participation in commercial AI: where implemented, participatory approaches in commercial AI labs are informal, atomised and often deprioritised, with limited incentive for companies to publicly declare adoption of participation approaches (even in the context of companies’ public commitments to fairness, trustworthiness, and other ethical principles). In some cases, interviewees confirmed concerns from the literature that participation-washing may be occurring. \par
Consequently, we conclude that factors such as the corporate profit motive and concern around exploitation are at present functioning as significant barriers to the use of participatory methods in AI , rather than drivers or enablers for the uptake of these practices. These concerns for the use of public participation in AI are exacerbated when one considers the growth of general purpose and generative AI systems, which enable a wide range of potential uses of AI systems in different contexts and settings. Successful public participation requires a clear use case for members of the public to understand, raising an innate challenge for the use of these methods for general purpose technologies. It is our intention for this research to function as a springboard: by presenting current conditions and emergent challenges for public participation in commercial AI, we lay foundations for further work and debate.

\section{Areas for further input}
The role of this paper is to provide insight into current challenges in public participation in commercial AI, but this is only one piece in the puzzle in better understanding the logics and conditions of participation in these environments. We acknowledge that possible next steps are manifold, require cooperation from multiple actors, and are unlikely to be ‘quick wins’. In light of some of our study limitations, further research on commercial AI public participation is necessary, such as ethnographic research of ‘live’ participatory projects in labs, to strengthen conclusions on the current lay of the land. \par
Second, the authors urge industry executives to exercise leadership in this area, namely: connect teams and individuals interested in ‘participatory AI’ across firms, provide institutional support and funding for further enquiry into participation in AI labs ‘in the open’ (with learnings made public), and vocally challenge the perceived norm of public participation working in opposition to tech business models. These combined forces may begin to unlock a grander normative vision for what participation in commercial AI should look like. \par
We join many of our interviewees in their demand for regulators and governments to incentivise this work through appropriate regulatory levers and offer funding and evaluation capacity to kickstart wider adoption of public participation. The authors also recognise and commend the contributions of activists, investigative journalists, researchers and others for their important work in raising awareness of tech industry abuses of power and in advancing algorithmic justice. We call on people affected by uses of AI, activists, civil society and other interest groups to maintain public pressure to advance a stake in the systems and technologies so often built using their data, but decoupled from their values, experiences and vision for technologies and society.
\section*{Acknowledgements}
We would like to thank Laura Carter, Emily Clough and Octavia Reeve for their review of this paper and for their thoughtful comments and suggestions. We have no grant sponsorship, external funding or conflicts of interest to declare.
\small\printbibliography

@article{moss_ethics_2020,
	title = {Ethics Owners: A New Model of Organizational Responsibility in Data-Driven Technology Companies},
	url = {https://datasociety.net/library/ethics-owners/},
	pages = {74},
	author = {Moss, Emanuel and Metcalf, Jacob},
	date = {2020},
	langid = {english},
}

@article{robertson_what_2020,
	title = {What If I Don't Like Any Of The Choices? The Limits of Preference Elicitation for Participatory Algorithm Design},
	url = {http://arxiv.org/abs/2007.06718},
	shorttitle = {What If I Don't Like Any Of The Choices?},
	journaltitle = {{arXiv}:2007.06718 [cs]},
	author = {Robertson, Samantha and Salehi, Niloufar},
	urldate = {2022-04-29},
	date = {2020-07-13},
	eprinttype = {arxiv},
	eprint = {2007.06718},
}

@article{martin_jr_participatory_2020,
	title = {Participatory Problem Formulation for Fairer Machine Learning Through Community Based System Dynamics},
	url = {http://arxiv.org/abs/2005.07572},
	journaltitle = {{arXiv}:2005.07572 [cs, stat]},
	author = {Martin Jr., Donald and Prabhakaran, Vinodkumar and Kuhlberg, Jill and Smart, Andrew and Isaac, William S.},
	urldate = {2022-04-13},
	date = {2020-05-22},
	eprinttype = {arxiv},
	eprint = {2005.07572},
}

@article{delgado_stakeholder_2021,
	title = {Stakeholder Participation in {AI}: Beyond "Add Diverse Stakeholders and Stir"},
	url = {http://arxiv.org/abs/2111.01122},
	shorttitle = {Stakeholder Participation in {AI}},
	author = {Delgado, Fernando and Yang, Stephen and Madaio, Michael and Yang, Qian},
	urldate = {2022-04-28},
	date = {2021-11-01},
}

@article{bratteteig_unpacking_2016,
	title = {Unpacking the Notion of Participation in Participatory Design},
	volume = {25},
	issn = {0925-9724, 1573-7551},
	url = {http://link.springer.com/10.1007/s10606-016-9259-4},
	doi = {10.1007/s10606-016-9259-4},
	pages = {425--475},
	number = {6},
	journaltitle = {Computer Supported Cooperative Work ({CSCW})},
	shortjournal = {Comput Supported Coop Work},
	author = {Bratteteig, Tone and Wagner, Ina},
	urldate = {2022-04-21},
	date = {2016-12},
	langid = {english},
}

@misc{birhane_power_2022,
	title = {Power to the People? Opportunities and Challenges for Participatory {AI}},
	url = {http://arxiv.org/abs/2209.07572},
	doi = {10.1145/3551624.3555290},
	shorttitle = {Power to the People?},
	author = {Birhane, Abeba and Isaac, William and Prabhakaran, Vinodkumar and Díaz, Mark and Elish, Madeleine Clare and Gabriel, Iason and Mohamed, Shakir},
	urldate = {2022-09-20},
	date = {2022-09-15},
	eprinttype = {arxiv},
	eprint = {2209.07572 [cs]},
}

@inproceedings{bondi_envisioning_2021,
	title = {Envisioning Communities: A Participatory Approach Towards {AI} for Social Good},
	url = {http://arxiv.org/abs/2105.01774},
	doi = {10.1145/3461702.3462612},
	shorttitle = {Envisioning Communities},
	pages = {425--436},
	booktitle = {Proceedings of the 2021 {AAAI}/{ACM} Conference on {AI}, Ethics, and Society},
	author = {Bondi, Elizabeth and Xu, Lily and Acosta-Navas, Diana and Killian, Jackson A.},
	urldate = {2022-09-19},
	date = {2021-07-21},
	eprinttype = {arxiv},
	eprint = {2105.01774 [cs]},
}

@inproceedings{barnett_crowdsourcing_2022,
	title = {Crowdsourcing Impacts: Exploring the Utility of Crowds for Anticipating Societal Impacts of Algorithmic Decision Making},
	url = {http://arxiv.org/abs/2207.09525},
	doi = {10.1145/3514094.3534145},
	shorttitle = {Crowdsourcing Impacts},
	pages = {56--67},
	booktitle = {Proceedings of the 2022 {AAAI}/{ACM} Conference on {AI}, Ethics, and Society},
	author = {Barnett, Julia and Diakopoulos, Nicholas},
	urldate = {2022-09-10},
	date = {2022-07-26},
	eprinttype = {arxiv},
	eprint = {2207.09525 [cs]},
}

@misc{hossain_towards_2021,
	title = {Towards a New Participatory Approach for Designing Artificial Intelligence and Data-Driven Technologies},
	url = {http://arxiv.org/abs/2104.04072},
	doi = {10.48550/arXiv.2104.04072},
	number = {{arXiv}:2104.04072},
	publisher = {{arXiv}},
	author = {Hossain, Soaad and Ahmed, Syed Ishtiaque},
	urldate = {2022-08-16},
	date = {2021-03-30},
	eprinttype = {arxiv},
	eprint = {2104.04072 [cs]},
}

@inproceedings{scott_algorithmic_2022,
	location = {Seoul Republic of Korea},
	title = {Algorithmic Tools in Public Employment Services: Towards a Jobseeker-Centric Perspective},
	isbn = {978-1-4503-9352-2},
	url = {https://dl.acm.org/doi/10.1145/3531146.3534631},
	doi = {10.1145/3531146.3534631},
	shorttitle = {Algorithmic Tools in Public Employment Services},
	eventtitle = {{FAccT} '22: 2022 {ACM} Conference on Fairness, Accountability, and Transparency},
	pages = {2138--2148},
	booktitle = {2022 {ACM} Conference on Fairness, Accountability, and Transparency},
	publisher = {{ACM}},
	author = {Scott, Kristen M. and Wang, Sonja Mei and Miceli, Milagros and Delobelle, Pieter and Sztandar-Sztanderska, Karolina and Berendt, Bettina},
	urldate = {2022-08-02},
	date = {2022-06-21},
	langid = {english},
}

@inproceedings{pierre_getting_2021,
	location = {Yokohama Japan},
	title = {Getting Ourselves Together: Data-centered participatory design research \& epistemic burden},
	isbn = {978-1-4503-8096-6},
	url = {https://dl.acm.org/doi/10.1145/3411764.3445103},
	doi = {10.1145/3411764.3445103},
	shorttitle = {Getting Ourselves Together},
	eventtitle = {{CHI} '21: {CHI} Conference on Human Factors in Computing Systems},
	pages = {1--11},
	booktitle = {Proceedings of the 2021 {CHI} Conference on Human Factors in Computing Systems},
	publisher = {{ACM}},
	author = {Pierre, Jennifer and Crooks, Roderic and Currie, Morgan and Paris, Britt and Pasquetto, Irene},
	urldate = {2022-06-29},
	date = {2021-05-06},
	langid = {english},
}

@report{patel_participatory_2021,
	title = {Participatory data stewardship},
	url = {https://www.adalovelaceinstitute.org/report/participatory-data-stewardship/},
	institution = {Ada Lovelace Institute},
	author = {Patel, Reema and Peppin, Aidan and Pavel, Valentina and Brennan, Jenny and Parker, Imogen and Safak, Jansu},
	urldate = {2022-01-10},
	date = {2021},
	langid = {british},
}

@book{costanza-chock_design_2020,
	location = {Cambridge, {MA}},
	title = {Design justice: community-led practices to build the worlds we need},
	isbn = {978-0-262-04345-8},
	series = {Information policy},
	shorttitle = {Design justice},
	publisher = {The {MIT} Press},
	author = {Costanza-Chock, Sasha},
	date = {2020},
}

@article{frey_this_2019,
	title = {"This Place Does What It Was Built For": Designing Digital Institutions for Participatory Change},
	volume = {3},
	url = {https://doi.org/10.1145/3359134},
	doi = {10.1145/3359134},
	shorttitle = {"This Place Does What It Was Built For"},
	pages = {32:1--32:31},
	issue = {{CSCW}},
	journaltitle = {Proceedings of the {ACM} on Human-Computer Interaction},
	shortjournal = {Proc. {ACM} Hum.-Comput. Interact.},
	author = {Frey, Seth and Krafft, P. M. and Keegan, Brian C.},
	urldate = {2022-10-14},
	date = {2019-11-07},
}

@misc{gilman_beyond_2022,
	location = {Rochester, {NY}},
	title = {Beyond Window Dressing: Public Participation for Marginalized Communities in the Datafied Society},
	url = {https://papers.ssrn.com/abstract=4266250},
	shorttitle = {Beyond Window Dressing},
	number = {4266250},
	author = {Gilman, Michele E.},
	urldate = {2022-11-17},
	date = {2022-11-02},
	langid = {english},
}

@misc{chan_limits_2021,
	title = {The Limits of Global Inclusion in {AI} Development},
	url = {http://arxiv.org/abs/2102.01265},
	number = {{arXiv}:2102.01265},
	publisher = {{arXiv}},
	author = {Chan, Alan and Okolo, Chinasa T. and Terner, Zachary and Wang, Angelina},
	urldate = {2023-01-11},
	date = {2021-02-01},
	eprinttype = {arxiv},
	eprint = {2102.01265 [cs]},
}

@article{frey_artificial_2020,
	title = {Artificial Intelligence and Inclusion: Formerly Gang-Involved Youth as Domain Experts for Analyzing Unstructured Twitter Data},
	volume = {38},
	issn = {0894-4393},
	url = {https://doi.org/10.1177/0894439318788314},
	doi = {10.1177/0894439318788314},
	shorttitle = {Artificial Intelligence and Inclusion},
	pages = {42--56},
	number = {1},
	journaltitle = {Social Science Computer Review},
	author = {Frey, William R. and Patton, Desmond U. and Gaskell, Michael B. and {McGregor}, Kyle A.},
	urldate = {2023-01-11},
	date = {2020-02-01},
	langid = {english},
	note = {Publisher: {SAGE} Publications Inc},
}

@article{cornwall_unpacking_2008,
	title = {Unpacking ‘Participation’: models, meanings and practices},
	volume = {43},
	issn = {0010-3802},
	url = {https://doi.org/10.1093/cdj/bsn010},
	doi = {10.1093/cdj/bsn010},
	shorttitle = {Unpacking ‘Participation’},
	pages = {269--283},
	number = {3},
	journaltitle = {Community Development Journal},
	shortjournal = {Community Development Journal},
	author = {Cornwall, Andrea},
	urldate = {2023-04-11},
	date = {2008-07-01},
}

@misc{seger_democratising_2023,
	title = {Democratising {AI}: Multiple Meanings, Goals, and Methods},
	url = {http://arxiv.org/abs/2303.12642},
	doi = {10.48550/arXiv.2303.12642},
	shorttitle = {Democratising {AI}},
	number = {{arXiv}:2303.12642},
	publisher = {{arXiv}},
	author = {Seger, Elizabeth and Ovadya, Aviv and Garfinkel, Ben and Siddarth, Divya and Dafoe, Allan},
	urldate = {2023-04-26},
	date = {2023-03-27},
	eprinttype = {arxiv},
	eprint = {2303.12642 [cs]},
}

@inproceedings{agid_its_2016,
	location = {New York, {NY}, {USA}},
	title = {``...it's your project, but it's not necessarily your work...": infrastructuring, situatedness, and designing relational practice},
	isbn = {978-1-4503-4046-5},
	url = {https://doi.org/10.1145/2940299.2940317},
	doi = {10.1145/2940299.2940317},
	series = {{PDC} '16},
	shorttitle = {"...it's your project, but it's not necessarily your work..."},
	pages = {81--90},
	booktitle = {Proceedings of the 14th Participatory Design Conference: Full papers - Volume 1},
	publisher = {Association for Computing Machinery},
	author = {Agid, Shana},
	urldate = {2022-04-21},
	date = {2016-08-15},
}

@article{arnstein_ladder_1969,
	title = {A Ladder Of Citizen Participation},
	volume = {35},
	issn = {0002-8991},
	url = {https://doi.org/10.1080/01944366908977225},
	doi = {10.1080/01944366908977225},
	pages = {216--224},
	number = {4},
	journaltitle = {Journal of the American Institute of Planners},
	author = {Arnstein, Sherry R.},
	urldate = {2022-05-11},
	date = {1969-07-01},
	note = {Publisher: Routledge
\_eprint: https://doi.org/10.1080/01944366908977225},
}

@online{atkinson_policymakers_2019,
	title = {A Policymaker’s Guide to the “Techlash”—What It Is and Why It’s a Threat to Growth and Progress {\textbar} {ITIF}},
	url = {https://itif.org/publications/2019/10/28/policymakers-guide-techlash/},
	author = {Atkinson, Robert and Brake, Doug and Castro, Daniel and Cunliff, Colin and Kennedy, Joe and {McLaughlin}, Michael and New, Joshua},
	urldate = {2023-05-10},
	date = {2019},
}

@online{dempsey_access_2021,
	title = {Access for all: the democratisation of {AI}},
	url = {https://eandt.theiet.org/content/articles/2021/11/access-for-all-the-democratisation-of-ai/},
	shorttitle = {Access for all},
	author = {Dempsey, Paul},
	urldate = {2023-02-01},
	date = {2021-11-10},
	langid = {american},
}

@article{cooper_accountability_2022,
	title = {Accountability in an Algorithmic Society: Relationality, Responsibility, and Robustness in Machine Learning},
	url = {http://arxiv.org/abs/2202.05338},
	doi = {10.1145/3531146.3533150},
	shorttitle = {Accountability in an Algorithmic Society},
	journaltitle = {{arXiv}:2202.05338 [cs]},
	author = {Cooper, A. Feder and Moss, Emanuel and Laufer, Benjamin and Nissenbaum, Helen},
	urldate = {2022-05-19},
	date = {2022-05-13},
	eprinttype = {arxiv},
	eprint = {2202.05338},
}

@article{himmelreich_against_2022,
	title = {Against “Democratizing {AI}”},
	issn = {0951-5666, 1435-5655},
	url = {https://link.springer.com/10.1007/s00146-021-01357-z},
	doi = {10.1007/s00146-021-01357-z},
	journaltitle = {{AI} \& {SOCIETY}},
	shortjournal = {{AI} \& Soc},
	author = {Himmelreich, Johannes},
	urldate = {2022-08-03},
	date = {2022-01-27},
	langid = {english},
}

@article{park_ai-based_2019,
	title = {{AI}-Based Request Augmentation to Increase Crowdsourcing Participation},
	volume = {7},
	rights = {Copyright (c) 2019 Association for the Advancement of Artificial Intelligence},
	issn = {2769-1349},
	url = {https://ojs.aaai.org/index.php/HCOMP/article/view/5282},
	doi = {10.1609/hcomp.v7i1.5282},
	pages = {115--124},
	journaltitle = {Proceedings of the {AAAI} Conference on Human Computation and Crowdsourcing},
	author = {Park, Junwon and Krishna, Ranjay and Khadpe, Pranav and Fei-Fei, Li and Bernstein, Michael},
	urldate = {2023-02-01},
	date = {2019-10-28},
	langid = {english},
}

@report{ada_lovelace_institute_algorithmic_2021,
	title = {Algorithmic accountability for the public sector},
	url = {https://www.opengovpartnership.org/documents/algorithmic-accountability-public-sector},
	pages = {70},
	institution = {Ada Lovelace Institute, {AI} Now Institute, Open Government Partnership},
	author = {Ada Lovelace Institute and {AI} Now Institute and Open Government Partnership},
	date = {2021},
	langid = {english},
}

@book{burr_introduction_2006,
	title = {An Introduction to Social Constructionism},
	isbn = {978-0-203-13302-6},
	url = {https://www.taylorfrancis.com/books/mono/10.4324/9780203133026/introduction-social-constructionism-vivien-burr},
	publisher = {Routledge},
	author = {Burr, Vivien},
	urldate = {2023-01-09},
	date = {2006-07-13},
	langid = {english},
	doi = {10.4324/9780203133026},
}

@article{bovens_analysing_2007-1,
	title = {Analysing and Assessing Public Accountability. A Conceptual Framework},
	url = {https://onlinelibrary.wiley.com/doi/abs/10.1111/j.1468-0386.2007.00378.x},
	pages = {37},
	journaltitle = {European Law Journal},
	author = {Bovens, Mark},
	date = {2007},
	langid = {english},
}

@online{openai_announcing_2023,
	title = {Announcing {OpenAI}’s Bug Bounty Program},
	url = {https://openai.com/blog/bug-bounty-program},
	author = {{OpenAI}},
	urldate = {2023-05-10},
	date = {2023},
}

@report{moss_assembling_2021,
	title = {Assembling Accountability: Algorithmic Impact Assessment for the Public Interest},
	url = {https://datasociety.net/library/assembling-accountability-algorithmic-impact-assessment-for-the-public-interest/},
	institution = {Data \& Society},
	author = {Moss, Emanuel and Watkins, Elizabeth Anne and Singh, Ranjit and Elish, Madeleine Clare and Metcalf, Jacob},
	date = {2021},
}

@book{eubanks_automating_2018,
	title = {Automating Inequality: How High-Tech Tools Profile, Police and Punish the Poor},
	author = {Eubanks, Virginia},
	date = {2018},
}

@book{habermas_between_1996,
	location = {Cambridge, {MA}, {USA}},
	title = {Between Facts and Norms: Contributions to a Discourse Theory of Law and Democracy},
	isbn = {978-0-262-08243-3},
	series = {Studies in Contemporary German Social Thought},
	shorttitle = {Between Facts and Norms},
	pagetotal = {676},
	publisher = {{MIT} Press},
	author = {Habermas, Jürgen},
	editorb = {{McCarthy}, Thomas},
	editorbtype = {redactor},
	translator = {Rehg, William},
	date = {1996-05-10},
	langid = {english},
}

@misc{denton_bringing_2020,
	title = {Bringing the People Back In: Contesting Benchmark Machine Learning Datasets},
	url = {http://arxiv.org/abs/2007.07399},
	shorttitle = {Bringing the People Back In},
	number = {{arXiv}:2007.07399},
	publisher = {{arXiv}},
	author = {Denton, Emily and Hanna, Alex and Amironesei, Razvan and Smart, Andrew and Nicole, Hilary and Scheuerman, Morgan Klaus},
	urldate = {2023-01-17},
	date = {2020-07-14},
	eprinttype = {arxiv},
	eprint = {2007.07399 [cs]},
}

@online{lloyd_camera_2020,
	title = {Camera Obscura: Beyond the lens of user-centered design},
	url = {https://alexis.medium.com/camera-obscura-beyond-the-lens-of-user-centered-design-631bb4f37594},
	shorttitle = {Camera Obscura},
	titleaddon = {Medium},
	author = {Lloyd, Alexis},
	urldate = {2023-01-11},
	date = {2020-12-21},
	langid = {english},
}

@online{iap2_core_nodate,
	title = {Core Values, Ethics, Spectrum – The 3 Pillars of Public Participation - International Association for Public Participation},
	url = {https://www.iap2.org/page/pillars},
	author = {{IAP2}},
	urldate = {2022-05-04},
}

@inproceedings{diaz_crowdworksheets_2022,
	title = {{CrowdWorkSheets}: Accounting for Individual and Collective Identities Underlying Crowdsourced Dataset Annotation},
	url = {http://arxiv.org/abs/2206.08931},
	doi = {10.1145/3531146.3534647},
	shorttitle = {{CrowdWorkSheets}},
	pages = {2342--2351},
	booktitle = {2022 {ACM} Conference on Fairness, Accountability, and Transparency},
	author = {Diaz, Mark and Kivlichan, Ian D. and Rosen, Rachel and Baker, Dylan K. and Amironesei, Razvan and Prabhakaran, Vinodkumar and Denton, Emily},
	urldate = {2023-02-01},
	date = {2022-06-21},
	eprinttype = {arxiv},
	eprint = {2206.08931 [cs]},
}

@article{mohamed_decolonial_2020,
	title = {Decolonial {AI}: Decolonial Theory as Sociotechnical Foresight in Artificial Intelligence},
	volume = {33},
	issn = {2210-5433, 2210-5441},
	url = {http://arxiv.org/abs/2007.04068},
	doi = {10.1007/s13347-020-00405-8},
	shorttitle = {Decolonial {AI}},
	pages = {659--684},
	number = {4},
	journaltitle = {Philosophy \& Technology},
	shortjournal = {Philos. Technol.},
	author = {Mohamed, Shakir and Png, Marie-Therese and Isaac, William},
	urldate = {2023-01-31},
	date = {2020-12},
	eprinttype = {arxiv},
	eprint = {2007.04068 [cs, stat]},
}

@article{bloomfield_deliberation_2001,
	title = {Deliberation and Inclusion: Vehicles for Increasing Trust in {UK} Public Governance?},
	volume = {19},
	issn = {0263-774X},
	url = {https://doi.org/10.1068/c6s},
	doi = {10.1068/c6s},
	shorttitle = {Deliberation and Inclusion},
	pages = {501--513},
	number = {4},
	journaltitle = {Environment and Planning C: Government and Policy},
	shortjournal = {Environ Plann C Gov Policy},
	author = {Bloomfield, Dan and Collins, Kevin and Fry, Charlotte and Munton, Richard},
	urldate = {2021-04-13},
	date = {2001-08-01},
	note = {Publisher: {SAGE} Publications Ltd {STM}},
}

@online{bit_deliberative_2022,
	title = {Deliberative democracy in action},
	url = {https://www.bi.team/blogs/deliberative-democracy-in-action/},
	author = {{BIT}},
	urldate = {2023-01-13},
	date = {2022},
	langid = {british},
}

@report{the_rsa_democratising_2019,
	title = {Democratising decisions about technology: a toolkit},
	url = {https://www.thersa.org/reports/democratising-decisions-technology-toolkit},
	shorttitle = {Democratising decisions about technology},
	author = {The {RSA}},
	urldate = {2023-02-03},
	date = {2019-10-24},
	langid = {english},
}

@article{stilgoe_developing_2013,
	title = {Developing a framework for responsible innovation},
	volume = {42},
	issn = {0048-7333},
	url = {https://www.sciencedirect.com/science/article/pii/S0048733313000930},
	doi = {10.1016/j.respol.2013.05.008},
	pages = {1568--1580},
	number = {9},
	journaltitle = {Research Policy},
	shortjournal = {Research Policy},
	author = {Stilgoe, Jack and Owen, Richard and Macnaghten, Phil},
	urldate = {2021-03-22},
	date = {2013-11-01},
	langid = {english},
}

@article{ocloo_exploring_2017,
	title = {Exploring the theory, barriers and enablers for patient and public involvement across health, social care and patient safety: a protocol for a systematic review of reviews},
	volume = {7},
	rights = {© Article author(s) (or their employer(s) unless otherwise stated in the text of the article) 2017. All rights reserved. No commercial use is permitted unless otherwise expressly granted.. This is an Open Access article distributed in accordance with the terms of the Creative Commons Attribution ({CC} {BY} 4.0) license, which permits others to distribute, remix, adapt and build upon this work, for commercial use, provided the original work is properly cited. See: http://creativecommons.org/licenses/by/4.0/},
	issn = {2044-6055, 2044-6055},
	url = {https://bmjopen.bmj.com/content/7/10/e018426},
	doi = {10.1136/bmjopen-2017-018426},
	shorttitle = {Exploring the theory, barriers and enablers for patient and public involvement across health, social care and patient safety},
	pages = {e018426},
	number = {10},
	journaltitle = {{BMJ} Open},
	author = {Ocloo, Josephine and Garfield, Sarah and Dawson, Shoba and Franklin, Bryony Dean},
	urldate = {2021-11-22},
	date = {2017-10-01},
	langid = {english},
	pmid = {29070642},
	note = {Publisher: British Medical Journal Publishing Group
Section: Health services research},
}

@inproceedings{bietti_ethics_2020,
	location = {New York, {NY}, {USA}},
	title = {From ethics washing to ethics bashing: a view on tech ethics from within moral philosophy},
	isbn = {978-1-4503-6936-7},
	url = {https://doi.org/10.1145/3351095.3372860},
	doi = {10.1145/3351095.3372860},
	series = {{FAT}* '20},
	shorttitle = {From ethics washing to ethics bashing},
	pages = {210--219},
	booktitle = {Proceedings of the 2020 Conference on Fairness, Accountability, and Transparency},
	publisher = {Association for Computing Machinery},
	author = {Bietti, Elettra},
	urldate = {2023-01-10},
	date = {2020-01-27},
}

@article{buolamwini_gender_2018,
	title = {Gender Shades: Intersectional Accuracy Disparities in Commercial Gender Classiﬁcation},
	pages = {15},
	author = {Buolamwini, Joy and Gebru, Timnit},
	date = {2018},
	langid = {english},
}

@article{vaughan_making_2017,
	title = {Making Better Use of the Crowd: How Crowdsourcing Can Advance Machine Learning Research},
	url = {https://jmlr.org/papers/v18/17-234.html},
	pages = {46},
	author = {Vaughan, Jennifer Wortman},
	date = {2017},
	langid = {english},
}

@report{berditchevskaia_participatory_2021,
	location = {London},
	title = {Participatory {AI} for humanitarian innovation: a briefing paper},
	url = {https://www.nesta.org.uk/report/participatory-ai-humanitarian-innovation-briefing-paper/},
	shorttitle = {Participatory {AI} for humanitarian innovation},
	institution = {Nesta},
	author = {Berditchevskaia, Aleks and Malliaraki, Eirini and Peach, Kathy},
	urldate = {2023-01-17},
	date = {2021},
	langid = {english},
}

@article{hugel_public_2020,
	title = {Public participation, engagement, and climate change adaptation: A review of the research literature},
	volume = {11},
	issn = {1757-7799},
	url = {https://onlinelibrary.wiley.com/doi/abs/10.1002/wcc.645},
	doi = {10.1002/wcc.645},
	shorttitle = {Public participation, engagement, and climate change adaptation},
	pages = {e645},
	number = {4},
	journaltitle = {{WIREs} Climate Change},
	author = {Hügel, Stephan and Davies, Anna R.},
	urldate = {2023-05-10},
	date = {2020},
	langid = {english},
	note = {\_eprint: https://onlinelibrary.wiley.com/doi/pdf/10.1002/wcc.645},
}

@article{banerjee_reading_2022,
	title = {Reading Race: {AI} Recognises Patient's Racial Identity In Medical Images},
	volume = {4},
	issn = {25897500},
	url = {http://arxiv.org/abs/2107.10356},
	doi = {10.1016/S2589-7500(22)00063-2},
	shorttitle = {Reading Race},
	pages = {e406--e414},
	number = {6},
	journaltitle = {The Lancet Digital Health},
	shortjournal = {The Lancet Digital Health},
	author = {Banerjee, Imon and Bhimireddy, Ananth Reddy and Burns, John L. and Celi, Leo Anthony and Chen, Li-Ching and Correa, Ramon and Dullerud, Natalie and Ghassemi, Marzyeh and Huang, Shih-Cheng and Kuo, Po-Chih and Lungren, Matthew P. and Palmer, Lyle and Price, Brandon J. and Purkayastha, Saptarshi and Pyrros, Ayis and Oakden-Rayner, Luke and Okechukwu, Chima and Seyyed-Kalantari, Laleh and Trivedi, Hari and Wang, Ryan and Zaiman, Zachary and Zhang, Haoran and Gichoya, Judy W.},
	urldate = {2023-01-13},
	date = {2022-06},
	eprinttype = {arxiv},
	eprint = {2107.10356 [cs, eess]},
}

@article{ganguli_red_2022,
	title = {Red Teaming Language Models to Reduce Harms: Methods, Scaling Behaviors, and Lessons Learned},
	author = {Ganguli, Deep and Lovitt, Liane and Kernion, Jackson and Askell, Amanda and Bai, Yuntao and Kadavath, Saurav and Mann, Ben and Perez, Ethan and Schiefer, Nicholas and Ndousse, Kamal and Jones, Andy and Bowman, Sam and Chen, Anna and Conerly, Tom and {DasSarma}, Nova and Drain, Dawn and Elhage, Nelson and El-Showk, Sheer and Fort, Stanislav and Dodds, Zac Hatfield and Henighan, Tom and Hernandez, Danny and Hume, Tristan and Jacobson, Josh and Johnston, Scott and Kravec, Shauna and Olsson, Catherine and Ringer, Sam and Tran-Johnson, Eli and Amodei, Dario and Brown, Tom and Joseph, Nicholas and {McCandlish}, Sam and Olah, Chris and Kaplan, Jared and Clark, Jack},
	date = {2022},
	langid = {english},
}

@online{dilhac_responsible_2021,
	title = {Responsible Artificial Intelligence: a Guide for Deliberation {\textbar} International observatory on the societal impacts of {AI} and digital technology},
	url = {https://observatoire-ia.ulaval.ca/en/responsible-artificial-intelligence-a-guide-for-deliberation/},
	author = {Dilhac, Marc-Antoine},
	urldate = {2023-05-10},
	date = {2021},
}

@online{hogarth_state_2022,
	title = {State of {AI} Report 2022},
	url = {https://www.stateof.ai/},
	author = {Hogarth, Nathan Benaich \{and\} Ian},
	urldate = {2023-02-02},
	date = {2022},
}

@inproceedings{boag_tech_2022,
	location = {Seoul Republic of Korea},
	title = {Tech Worker Organizing for Power and Accountability},
	isbn = {978-1-4503-9352-2},
	url = {https://dl.acm.org/doi/10.1145/3531146.3533111},
	doi = {10.1145/3531146.3533111},
	eventtitle = {{FAccT} '22: 2022 {ACM} Conference on Fairness, Accountability, and Transparency},
	pages = {452--463},
	booktitle = {2022 {ACM} Conference on Fairness, Accountability, and Transparency},
	publisher = {{ACM}},
	author = {Boag, William and Suresh, Harini and Lepe, Bianca and D'Ignazio, Catherine},
	urldate = {2022-06-30},
	date = {2022-06-21},
	langid = {english},
}

@article{kochan_commenting_2017,
	title = {The Commenting Power: Agency Accountability through Public Participation},
	issn = {1556-5068},
	url = {https://www.ssrn.com/abstract=3006157},
	doi = {10.2139/ssrn.3006157},
	shorttitle = {The Commenting Power},
	journaltitle = {{SSRN} Electronic Journal},
	shortjournal = {{SSRN} Journal},
	author = {Kochan, Donald J.},
	urldate = {2021-03-19},
	date = {2017},
	langid = {english},
}

@article{dryzek_crisis_2019,
	title = {The crisis of democracy and the science of deliberation},
	volume = {363},
	url = {https://www.science.org/doi/10.1126/science.aaw2694},
	doi = {10.1126/science.aaw2694},
	pages = {1144--1146},
	number = {6432},
	journaltitle = {Science},
	author = {Dryzek, John S. and Bächtiger, André and Chambers, Simone and Cohen, Joshua and Druckman, James N. and Felicetti, Andrea and Fishkin, James S. and Farrell, David M. and Fung, Archon and Gutmann, Amy and Landemore, Hélène and Mansbridge, Jane and Marien, Sofie and Neblo, Michael A. and Niemeyer, Simon and Setälä, Maija and Slothuus, Rune and Suiter, Jane and Thompson, Dennis and Warren, Mark E.},
	urldate = {2023-05-10},
	date = {2019-03-15},
	note = {Publisher: American Association for the Advancement of Science},
}

@article{couldry_decolonial_2021,
	title = {The decolonial turn in data and technology research: what is at stake and where is it heading?},
	volume = {0},
	issn = {1369-118X},
	url = {https://doi.org/10.1080/1369118X.2021.1986102},
	doi = {10.1080/1369118X.2021.1986102},
	shorttitle = {The decolonial turn in data and technology research},
	pages = {1--17},
	number = {0},
	journaltitle = {Information, Communication \& Society},
	author = {Couldry, Nick and Mejias, Ulises Ali},
	urldate = {2023-01-23},
	date = {2021-11-09},
	note = {Publisher: Routledge
\_eprint: https://doi.org/10.1080/1369118X.2021.1986102},
}

@article{harrington_forgotten_2020,
	title = {The forgotten margins: what is community-based participatory health design telling us?},
	volume = {27},
	issn = {1072-5520, 1558-3449},
	url = {https://dl.acm.org/doi/10.1145/3386381},
	doi = {10.1145/3386381},
	shorttitle = {The forgotten margins},
	pages = {24--29},
	number = {3},
	journaltitle = {Interactions},
	shortjournal = {interactions},
	author = {Harrington, Christina N.},
	urldate = {2023-01-11},
	date = {2020-04-17},
	langid = {english},
}

@article{dimaggio_iron_1983,
	title = {The Iron Cage Revisited: Institutional Isomorphism and Collective Rationality in Organizational Fields},
	volume = {48},
	issn = {0003-1224},
	url = {https://www.jstor.org/stable/2095101},
	doi = {10.2307/2095101},
	shorttitle = {The Iron Cage Revisited},
	pages = {147--160},
	number = {2},
	journaltitle = {American Sociological Review},
	author = {{DiMaggio}, Paul J. and Powell, Walter W.},
	urldate = {2023-01-09},
	date = {1983},
	note = {Publisher: [American Sociological Association, Sage Publications, Inc.]},
}

@misc{birhane_values_2022,
	title = {The Values Encoded in Machine Learning Research},
	url = {http://arxiv.org/abs/2106.15590},
	number = {{arXiv}:2106.15590},
	publisher = {{arXiv}},
	author = {Birhane, Abeba and Kalluri, Pratyusha and Card, Dallas and Agnew, William and Dotan, Ravit and Bao, Michelle},
	urldate = {2023-01-07},
	date = {2022-06-21},
	eprinttype = {arxiv},
	eprint = {2106.15590 [cs]},
}

@inproceedings{katell_toward_2020,
	location = {New York, {NY}, {USA}},
	title = {Toward situated interventions for algorithmic equity: lessons from the field},
	isbn = {978-1-4503-6936-7},
	url = {https://dl.acm.org/doi/10.1145/3351095.3372874},
	doi = {10.1145/3351095.3372874},
	series = {{FAT}* '20},
	shorttitle = {Toward situated interventions for algorithmic equity},
	pages = {45--55},
	booktitle = {Proceedings of the 2020 Conference on Fairness, Accountability, and Transparency},
	publisher = {Association for Computing Machinery},
	author = {Katell, Michael and Young, Meg and Dailey, Dharma and Herman, Bernease and Guetler, Vivian and Tam, Aaron and Bintz, Corinne and Raz, Daniella and Krafft, P. M.},
	urldate = {2023-05-10},
	date = {2020-01-27},
}

@inproceedings{suresh_towards_2022,
	location = {Seoul Republic of Korea},
	title = {Towards Intersectional Feminist and Participatory {ML}: A Case Study in Supporting Feminicide Counterdata Collection},
	isbn = {978-1-4503-9352-2},
	url = {https://dl.acm.org/doi/10.1145/3531146.3533132},
	doi = {10.1145/3531146.3533132},
	shorttitle = {Towards Intersectional Feminist and Participatory {ML}},
	eventtitle = {{FAccT} '22: 2022 {ACM} Conference on Fairness, Accountability, and Transparency},
	pages = {667--678},
	booktitle = {2022 {ACM} Conference on Fairness, Accountability, and Transparency},
	publisher = {{ACM}},
	author = {Suresh, Harini and Movva, Rajiv and Dogan, Amelia Lee and Bhargava, Rahul and Cruxen, Isadora and Cuba, Angeles Martinez and Taurino, Guilia and So, Wonyoung and D'Ignazio, Catherine},
	urldate = {2022-06-29},
	date = {2022-06-21},
	langid = {english},
}

@misc{cobbe_understanding_2023,
	location = {Rochester, {NY}},
	title = {Understanding accountability in algorithmic supply chains},
	url = {https://papers.ssrn.com/abstract=4430778},
	number = {4430778},
	author = {Cobbe, Jennifer and Veale, Michael and Singh, Jatinder},
	urldate = {2023-05-04},
	date = {2023-04-07},
	langid = {english},
}

@article{brodie_understanding_2009,
	title = {Understanding participation:},
	pages = {50},
	journaltitle = {An introduction},
	author = {Brodie, Ellie and Cowling, Eddie and Nissen, Nina},
	date = {2009},
	langid = {english},
}

@article{braun_using_2006,
	title = {Using thematic analysis in psychology},
	volume = {3},
	issn = {1478-0887, 1478-0895},
	url = {http://www.tandfonline.com/doi/abs/10.1191/1478088706qp063oa},
	doi = {10.1191/1478088706qp063oa},
	pages = {77--101},
	number = {2},
	journaltitle = {Qualitative Research in Psychology},
	shortjournal = {Qualitative Research in Psychology},
	author = {Braun, Virginia and Clarke, Victoria},
	urldate = {2022-11-08},
	date = {2006-01},
	langid = {english},
}

@inproceedings{dove_ux_2017,
	location = {New York, {NY}, {USA}},
	title = {{UX} Design Innovation: Challenges for Working with Machine Learning as a Design Material},
	isbn = {978-1-4503-4655-9},
	url = {https://doi.org/10.1145/3025453.3025739},
	doi = {10.1145/3025453.3025739},
	series = {{CHI} '17},
	shorttitle = {{UX} Design Innovation},
	pages = {278--288},
	booktitle = {Proceedings of the 2017 {CHI} Conference on Human Factors in Computing Systems},
	publisher = {Association for Computing Machinery},
	author = {Dove, Graham and Halskov, Kim and Forlizzi, Jodi and Zimmerman, John},
	urldate = {2023-01-23},
	date = {2017-05-02},
}

@article{flanders_what_2013,
	title = {What Is the Value of Participation?},
	volume = {66},
	journaltitle = {{OKLAHOMA} {LAW} {REVIEW}},
	author = {Flanders, Chad W},
	date = {2013},
	langid = {english},
}

@article{rakova_where_2021,
	title = {Where Responsible {AI} meets Reality: Practitioner Perspectives on Enablers for shifting Organizational Practices},
	volume = {5},
	issn = {2573-0142},
	url = {http://arxiv.org/abs/2006.12358},
	doi = {10.1145/3449081},
	shorttitle = {Where Responsible {AI} meets Reality},
	pages = {1--23},
	issue = {{CSCW}1},
	journaltitle = {Proceedings of the {ACM} on Human-Computer Interaction},
	shortjournal = {Proc. {ACM} Hum.-Comput. Interact.},
	author = {Rakova, Bogdana and Yang, Jingying and Cramer, Henriette and Chowdhury, Rumman},
	urldate = {2023-04-11},
	date = {2021-04-13},
	eprinttype = {arxiv},
	eprint = {2006.12358 [cs]},
}

@article{glimmerveen_who_2022,
	title = {Who Participates in Public Participation? The Exclusionary Effects of Inclusionary Efforts},
	volume = {54},
	issn = {0095-3997},
	url = {https://doi.org/10.1177/00953997211034137},
	doi = {10.1177/00953997211034137},
	shorttitle = {Who Participates in Public Participation?},
	pages = {543--574},
	number = {4},
	journaltitle = {Administration \& Society},
	shortjournal = {Administration \& Society},
	author = {Glimmerveen, Ludo and Ybema, Sierk and Nies, Henk},
	urldate = {2022-05-12},
	date = {2022-04-01},
	langid = {english},
	note = {Publisher: {SAGE} Publications Inc},
}

@article{kalluri_dont_2020-1,
	title = {Don’t ask if artificial intelligence is good or fair, ask how it shifts power},
	volume = {583},
	url = {https://www.nature.com/articles/d41586-020-02003-2},
	doi = {10.1038/d41586-020-02003-2},
	pages = {169--169},
	number = {7815},
	journaltitle = {Nature},
	author = {Kalluri, Pratyusha},
	urldate = {2022-05-16},
	date = {2020-07-07},
}

@article{solomon_why_2012-1,
	title = {Why and When Should We Use Public Deliberation?},
	volume = {42},
	issn = {1552-146X},
	url = {https://onlinelibrary.wiley.com/doi/abs/10.1002/hast.27},
	doi = {10.1002/hast.27},
	pages = {17--20},
	number = {2},
	journaltitle = {Hastings Center Report},
	author = {Solomon, Stephanie and Abelson, Julia},
	urldate = {2022-05-12},
	date = {2012},
}

@article{cleaver_paradoxes_1999-1,
	title = {Paradoxes of participation: questioning participatory approaches to development},
	volume = {11},
	issn = {1099-1328},
	url = {https://onlinelibrary.wiley.com/doi/abs/10.1002/%28SICI%291099-1328%28199906%2911%3A4%3C597%3A%3AAID-JID610%3E3.0.CO%3B2-Q},
	doi = {10.1002/(SICI)1099-1328(199906)11:4<597::AID-JID610>3.0.CO;2-Q},
	shorttitle = {Paradoxes of participation},
	pages = {597--612},
	number = {4},
	journaltitle = {Journal of International Development},
	author = {Cleaver, Frances},
	urldate = {2023-01-25},
	date = {1999},
}

@online{probiner_smart_2018,
	title = {From smart products to smart systems},
	url = {https://www2.deloitte.com/content/www/us/en/insights/focus/cognitive-technologies/participatory-design-artificial-intelligence.html},
	author = {Probiner, Scott and Murphy, Timothy},
	urldate = {2023-01-31},
	date = {2018},
}

@article{christiano_deep_2017,
	title = {Deep reinforcement learning from human preferences},
	rights = {{arXiv}.org perpetual, non-exclusive license},
	url = {https://arxiv.org/abs/1706.03741},
	doi = {10.48550/ARXIV.1706.03741},
	author = {Christiano, Paul and Leike, Jan and Brown, Tom B. and Martic, Miljan and Legg, Shane and Amodei, Dario},
	urldate = {2023-05-11},
	date = {2017},
	note = {Publisher: {arXiv}
Version Number: 4},
}

@article{glucker_public_2013,
	title = {Public participation in environmental impact assessment: why, who and how?},
	volume = {43},
	issn = {01959255},
	url = {https://linkinghub.elsevier.com/retrieve/pii/S0195925513000711},
	doi = {10.1016/j.eiar.2013.06.003},
	shorttitle = {Public participation in environmental impact assessment},
	pages = {104--111},
	journaltitle = {Environmental Impact Assessment Review},
	shortjournal = {Environmental Impact Assessment Review},
	author = {Glucker, Anne N. and Driessen, Peter P.J. and Kolhoff, Arend and Runhaar, Hens A.C.},
	urldate = {2023-05-11},
	date = {2013-11},
	langid = {english},
}

@article{holstein_co-designing_2019,
	title = {Co-Designing a Real-Time Classroom Orchestration Tool to Support Teacher–{AI} Complementarity},
	volume = {6},
	issn = {1929-7750},
	url = {https://learning-analytics.info/index.php/JLA/article/view/6336},
	doi = {10.18608/jla.2019.62.3},
	number = {2},
	journaltitle = {Journal of Learning Analytics},
	shortjournal = {Learning Analytics},
	author = {Holstein, Kenneth and {McLaren}, Bruce M. and Aleven, Vincent},
	urldate = {2023-05-11},
	date = {2019-07-22},
}

@online{openai_how_2023,
	title = {How should {AI} systems behave, and who should decide?},
	url = {https://openai.com/blog/how-should-ai-systems-behave},
	author = {{OpenAI}},
	urldate = {2023-04-20},
	date = {2023-02-17},
}

@article{russell_impact_2020,
	title = {The impact of public involvement in health research: what are we measuring? Why are we measuring it? Should we stop measuring it?},
	volume = {6},
	issn = {2056-7529},
	url = {https://researchinvolvement.biomedcentral.com/articles/10.1186/s40900-020-00239-w},
	doi = {10.1186/s40900-020-00239-w},
	shorttitle = {The impact of public involvement in health research},
	pages = {63},
	number = {1},
	journaltitle = {Research Involvement and Engagement},
	shortjournal = {Res Involv Engagem},
	author = {Russell, Jill and Fudge, Nina and Greenhalgh, Trish},
	urldate = {2023-05-11},
	date = {2020-12},
	langid = {english},
}

@article{sloane_make_2022,
	title = {To make {AI} fair, here’s what we must learn to do},
	volume = {605},
	issn = {0028-0836, 1476-4687},
	url = {https://www.nature.com/articles/d41586-022-01202-3},
	doi = {10.1038/d41586-022-01202-3},
	pages = {9--9},
	number = {7908},
	journaltitle = {Nature},
	shortjournal = {Nature},
	author = {Sloane, Mona},
	urldate = {2023-05-11},
	date = {2022-05-05},
	langid = {english},
}

@article{saetra_avoid_2022,
	title = {Avoid diluting democracy by algorithms},
	volume = {4},
	issn = {2522-5839},
	url = {https://www.nature.com/articles/s42256-022-00537-w},
	doi = {10.1038/s42256-022-00537-w},
	pages = {804--806},
	number = {10},
	journaltitle = {Nature Machine Intelligence},
	shortjournal = {Nat Mach Intell},
	author = {Sætra, Henrik Skaug and Borgebund, Harald and Coeckelbergh, Mark},
	urldate = {2023-05-11},
	date = {2022-09-29},
	langid = {english},
}

@misc{sloane_participation_2020,
	title = {Participation is not a Design Fix for Machine Learning},
	url = {http://arxiv.org/abs/2007.02423},
	number = {{arXiv}:2007.02423},
	publisher = {{arXiv}},
	author = {Sloane, Mona and Moss, Emanuel and Awomolo, Olaitan and Forlano, Laura},
	urldate = {2023-05-11},
	date = {2020-08-11},
	eprinttype = {arxiv},
	eprint = {2007.02423 [cs]},
}

@collection{jacko_human-computer_2003,
	location = {Mahwah, N.J},
	title = {The human-computer interaction handbook: fundamentals, evolving technologies, and emerging applications},
	isbn = {978-0-429-16397-5},
	series = {Human factors and ergonomics},
	shorttitle = {The human-computer interaction handbook},
	pagetotal = {1277},
	publisher = {Lawrence Erlbaum Associates},
	editor = {Jacko, Julie A. and Sears, Andrew},
	date = {2003},
}
}
\end{multicols}

\end{document}